\documentclass[fleqn,usenatbib]{mnras}

\usepackage{newtxtext,newtxmath}

\usepackage[T1]{fontenc}

\DeclareRobustCommand{\VAN}[3]{#2}
\let\VANthebibliography\thebibliography
\def\thebibliography{\DeclareRobustCommand{\VAN}[3]{##3}\VANthebibliography}

\usepackage{graphicx}	
\usepackage{amsmath}	

\newcommand*  {\diff}       {\mathop{}\!\mathrm{d}}
\newcommand*  {\p}          {\partial}
\renewcommand*{\vec}[1]     {\boldsymbol{#1}}
\newcommand*  {\uvec}[1]	{\hat{\vec{#1}}}
\newcommand*  {\tens}[1]    {\vec{\mathsf{#1}}}
\newcommand*  {\I}      	{\mathrm{i}}
\newcommand*  {\Exp}[1]     {\mathrm{e}^{#1}}
\newcommand*  {\tdiff}[2]   {\frac{\diff{#1}}{\diff{#2}}}
\newcommand*  {\pdiff}[2]   {\frac{\p{#1}}{\p{#2}}}
\newcommand*  {\cross}	    {\times}
\newcommand*  {\sub}[2]     {{#1}_{\mathrm{#2}}}
\newcommand   {\new}[1]     {#1}

\title[Measuring pattern speeds of simulated bars]{Measuring bar pattern speeds from single simulation snapshots}

\author[Dehnen, Semczuk \& Sch{\"o}nrich]{%
\href{http://orcid.org/0000-0001-8669-2316}{Walter Dehnen\includegraphics[width=11pt]{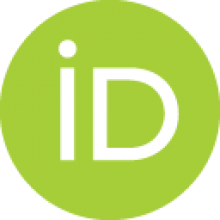}},$^{\!\!1,2,3}$
\href{http://orcid.org/0000-0002-8191-8918}{Marcin Semczuk\includegraphics[width=11pt]{orcid-ID}},$^{\!\!2}$ and
\href{http://orcid.org/0000-0002-4236-3091}{Ralph Sch{\"o}nrich\includegraphics[width=11pt]{orcid-ID}},$^{\!\!4}$
\smallskip
\\
$^1$ Astronomisches Recheninstitut, Zentrum f{\"u}r Astronomie der Universit{\"a}t Heidelberg, M{\"o}nchhofstra\ss{}e 12-14, 69120, Heidelberg, Germany \\
$^2$ School for Physics and Astronomy, University of Leicester, University Road, LE1 7RH, Leicester, UK \\
$^3$ Universit{\"a}ts-Sternwarte M{\"u}nchen, Scheinerstra\ss{}e 1, 81679, M{\"u}nchen, Germany\\
$^4$ Mullard Space Science Laboratory, University College London, Holmbury St.~Mary, Dorking, Surrey, RH5 6NT, UK
}

\date{Accepted XXX. Received YYY; in original form ZZZ}

\pubyear{2021}

\begin{document}
\label{firstpage}
\pagerange{\pageref{firstpage}--\pageref{lastpage}}
\maketitle

\begin{abstract}
We describe methods to measure simultaneously the orientation angle $\psi$ and pattern speed $\Omega
$ from single snapshots of simulated barred galaxies. Unlike previous attempts, our approach is unbiased\new{, precise, and consistent in the sense that $\psi=\int\Omega\diff t$}. It can be extended to obtain the rate \emph{and} axis of rotation, i.e.\ the vector $\vec{\Omega}$. We provide computer code implementing our method.
\end{abstract}


\maketitle

\section{Introduction}
About two thirds of spiral galaxies in the local universe host a central stellar bar \citep{EskridgeEtal2000, MenendezDelmestreEtAl2007, ShethEtAl2008, Masters2011, Cheung2013, Erwin2018}. Such bars are thought to rotate almost rigidly with an angular frequency or pattern speed $\Omega$. In simulations $\Omega$ is commonly found to slowly decrease with time, as angular momentum is transferred from the bar to the dark halo, which is generally accompanied by a growth in bar strength and length \citep{Sellwood1980, Weinberg1985, LittleCarlberg1991, DebattistaSellwood1998, Athanassoula2003}. The rate of slowdown depends on the mass and structure of the bar, and on the balance between angular the momentum absorbed by the halo and that surrendered by gas driven into the galactic centre by the bar. Recently, \cite{Chiba2021} found in the stellar kinematics of the Solar neighbourhood evidence for the slowdown of the Milky Way bar.

Overlaying this continuous slowdown, bar pattern speeds (along with other bar parameters, see \citealt{Wu2016}) are subject to short-term oscillations, e.g. by bar-spiral interactions  \citep{WuPfennigerTaam2018, Hilmi2020}.

A dimensionless parameter for the rotation of a bar is the ratio $\mathcal{R}=R_\mathrm{CR}/R_\mathrm{bar}$ between the radius $R_\mathrm{CR}$ at which a star on a circular orbit co-rotates with the bar, and the bar's actual size $R_\mathrm{bar}$. The confinement of most bar-supporting orbits to $R<R_\mathrm{CR}$, sets the theoretical limit $\mathcal{R}\ge1$. Observational determinations of both $R_\mathrm{CR}$ and $R_\mathrm{bar}$ are plagued with difficulties and systematic uncertainties, but largely suggest that bars rotate nearly as fast as possible, i.e.\ $\mathcal{R}\lesssim1.4$ \citep{Corsini2011, Aguerri2015, Guo2019}. Simulations of galaxy formation, on the other hand, tend to predict bars to be slower \citep{Algorry2017, Peschken2019} or shorter \citep[][using the IllustrisTNG simulation]{Frankel2022}. \cite{Fragkoudi2021} suggest that this tension lessens when increasing the resolution of the models (in other words, the models may not yet be converged on $\mathcal{R}$).

A deeper understanding of bar slowdown, pattern speed oscillations, expected distribution of $\mathcal{R}$ and other bar rotation-related topics all depend on accurate measurements of $\Omega(t)$ from simulations. In simulations with high output cadence, the most common method is to derive the bar angle $\psi$ from an $m=2$ Fourier analysis \citep{Sellwood1986} of consecutive snapshots and calculate the pattern speed as finite difference:
\begin{align}
    \label{eq:Om:FD}
    \Omega \approx \Omega_{\mathrm{FD}} \equiv \frac{\Delta\psi}{\Delta t}.
\end{align}
Since $\psi$ is $\upi$-periodic, this simple method requires $\Omega\,\Delta t\ll\upi$ to unambiguously identify $\Delta\psi$, and is therefore not viable for simulations with long output intervals $\Delta t$ (or if $\Omega$ is required on the fly during a simulation). This is the typical situation for large cosmological simulations, when data volume limits the output frequency. 

In this situation, \cite{Peschken2019} and \cite{Fragkoudi2021} applied the \cite{TremaineWeinberg1984} method for determining $\Omega$ of external galaxies from line-of-sight velocities. They report an accuracy of $\sim10\%$ and $\sim 5\,$km\,s$^{-1}$kpc$^{-1}$ respectively, depending also on the adopted viewing angle. Applying the Tremaine-Weinberg method to simulations may be justified for direct comparison to observations, but is certainly not ideal. This is because it relies on the assumption of stationarity of the pattern (which is generally not satisfied as mentioned above) and utilises only one of three Cartesian velocity components.

The orientation $\psi$ of simulated bars is well measured from the particle positions $\vec{x}_i$ (and their masses) as phase of the $m=2$ Fourier component. Since the particles move, $\psi$ is an implicit function of time $t$, which can be differentiated to obtain
\begin{align}
    \Omega &= \frac{\diff\psi}{\diff t} = \sum_i \frac{\partial\psi}{\partial\vec{x}_i} \cdot \tdiff{\vec{x}_i}{t} .
\end{align}
Crucially, $\psi$ depends on $\vec{x}_i$ not only through the azimuth $\varphi_i$, which enters the Fourier analysis, but also through the spatial selection, usually from an annulus. \new{Neglecting this dependence \citep{WuPfennigerTaam2018} ignores the difference between the particle sets from which $\psi$ is measured at $t$ and $t+\Delta t$ and results in systematic errors of 5-25\% (see Fig.~\ref{fig:test0}).}

\new{When using radial bins (annuli), their sharp boundaries generate divergent $\p\psi/\p R_i$, which cannot be evaluated for particle systems. Instead, \cite{Frankel2022} estimated the resulting dependence on the particle velocities in a not reproducibly specified way and report that the accuracy for $\Omega$ seems to be $\sim10$\%. However, such a treatment cannot be consistent in the sense that $\psi$ and $\Omega$ measured from the particles satisfy $\psi=\int\Omega\diff t$.}

\new{Therefore, for a consistent measurement of $\Omega$} annuli with sharp boundaries must be avoided in favour of smoothly varying window functions. This is analogous to the way local properties are estimated in smoothed-particle hydrodynamics (SPH).

We also show that the $m=2$ Fourier method for identifying $\psi$ is equivalent to obtaining $\psi$ as the direction of the eigenvectors of the moment of inertia. This insight provides a way to measure the pattern speed vector $\vec{\Omega}$, i.e.\ no longer \emph{assuming} rotation around a particular axis. This generalised method may be suitable to measure the tumbling rate and axis of simulated triaxial galaxies and haloes.

This paper is organised as follows. Section~\ref{sec:method} contains the derivation of the Fourier and moment-of-inertia methods, Section~\ref{sec:test} describes tests of the method on a suite of $N$-body simulations and Section~\ref{sec:discussion} summarises and discusses our findings.

\section{Methods}
\label{sec:method}
Before measuring a pattern speed, the centre of rotation must be known. Here we do not discuss finding the centre, as various good methods have been published (e.g.\ the shrinking sphere, \citealt{Power2003}), but note that also the rate of change of the centre position (which may differ from the central velocity) must be known with uncertainty well below the velocity dispersion. In the remainder, $\vec{x}$ and $\vec{v}$ denote position and velocity relative to that centre.

We begin by assuming that rotation is around the $z$-axis, i.e. in azimuth $\varphi$, and that the density $\rho$ is stationary in a frame rotating with angular rate $\Omega(t)$, but will later relax both assumptions. With these assumptions, $\rho(R,z,\varphi,t)=f(R,z,\varphi-\psi)$ with some function $f$ and the instantaneous orientation 
\begin{align}
    \label{eq:psi=intOmega}
    \textstyle \psi(t) \equiv \int\Omega(t)\diff t
\end{align}
of the rotating frame, such that $\p\rho/\p t=-\Omega\,\p\rho/\p\varphi$. Combining this with the continuity equation $\p\rho/\p t+\vec{\nabla}\cdot(\rho\bar{\vec{v}})=0$, we find
\begin{align}
	\label{eq:cont:3D:z}
    \Omega \pdiff{\rho}{\varphi} = \vec{\nabla}\cdot(\rho\bar{\vec{v}}),
\end{align}
where $\bar{\vec{v}}(\vec{x})$ is the mean (streaming) velocity. We exploit equation~\eqref{eq:cont:3D:z} by multiplying both sides by a weight function $w(\vec{x})$ and integrating over all space to find
\begin{align}
	\label{eq:TW}
	\Omega \int\!\!\diff^3\!\vec{x}\,\rho\,\pdiff{w}{\varphi}
    = \int\!\!\diff^3\!\vec{x}\,\rho\bar{\vec{v}}\cdot\vec{\nabla}w,
\end{align}
where we have integrated each side by parts to shift the derivatives onto $w$. The \cite{TremaineWeinberg1984} method is obtained from equation~\eqref{eq:TW} by weighing with the Heaviside function, $w=\Theta(y-y_0)$, which reduces the velocity term to its $y$ component and the integrals to a slit at $y=y_0$.

\subsection{Fourier methods}
\label{sec:fourier}
A natural choice for the weight function is $w(\vec{x})=W(R)\,\Exp{-\I m\varphi}$, where $m$ is an azimuthal wave\-number and $W\ge0$ some window function. Equation~\eqref{eq:TW} then yields
\begin{align}
	\label{eq:O:Wf}
	\Omega = \frac{\int\diff^3\!\vec{x}\,\rho \left[W \bar{\dot{\varphi}} + \frac{\I}{m} \bar{v}_R(\p W/\p R)\right]\Exp{-\I m\varphi}}{\int\diff^3\!\vec{x}\,\rho\,W\,\Exp{-\I m\varphi}}.
\end{align}
Since this equation was derived under the assumption that $\rho$ is stationary in the rotating frame, the right-hand side is real valued, provided this assumption is satisfied. However, since bars often evolve, we now relax this assumption, when the right-hand side of equation~\eqref{eq:O:Wf} generally includes an imaginary part. In this case, the real part remains the correct answer for $\Omega$. To show this, we define the window-averaged surface density $\Sigma(\varphi,t)\equiv\iint \rho W R\diff R\diff z$ and take its azimuthal Fourier transform
\begin{align}
	\label{eq:Four:trans}
	\hat{\Sigma}_{m}(t)
	    &= \frac{1}{2\upi}\int\diff^3\!\vec{x}\,W(R)\,\rho(\vec{x},t)\,\Exp{-\I m\varphi}
\end{align}
with time derivative
\begin{align}
	\label{eq:rho:m|w}
	\tdiff{\hat{\Sigma}_{m}}{t}
	&=\frac{1}{2\upi}\int\diff^3\!\vec{x}\,\rho\;
		\left[- \I m \bar{\dot{\varphi}} W + \bar{v}_R\pdiff{W}{R}\right]\,\Exp{-\I m\varphi},
\end{align}
where we used the continuity equation to eliminate $\p\rho/\p t$. Expressing $\hat{\Sigma}_{m}$ in polar form with amplitude $\Sigma_{m}$ and phase $\psi_{m}$, and identifying $\Omega=\dot{\psi}_{m}$ gives with equation~\eqref{eq:rho:m|w}
\begin{align}
	\label{eq:Omega:Fourier}
	\Omega + \frac{\I}{m} \frac{\dot\Sigma_{m}}{\Sigma_m}
	&= \frac
    {\int\diff^3\!\vec{x}\,\rho \left[\bar{\dot{\varphi}} W + \frac{\I}{m} \bar{v}_R(\p W/\p R)\right]\,\Exp{-\I m\varphi}}
	{\int\diff^3\!\vec{x}\,\rho\,W\,\Exp{-\I m\varphi}}
\end{align}
with right-hand side identical to equation~\eqref{eq:O:Wf}. Thus, only the real part of these right-hand sides measures a pattern speed, namely that of the azimuthal $m$-wave in the window $W$, while the imaginary part is related to the rate of change in wave amplitude $\Sigma_m$. 

Finally, for application to $N$-body models, we transform equation~\eqref{eq:Omega:Fourier} via the usual substitutions $\int\diff^3\!\vec{x}\rho \to \sum_i\mu_i$ and $\int\diff^3\!\vec{x}\rho\bar{\vec{v}} \to \sum_i\mu_i\vec{v}_i$, with particle masses $\mu_i$, to its discrete form
\begin{align}
	\label{eq:Omega:Fourier:i}
	\Omega + \frac{\I}{m} \frac{\dot\Sigma_{m}}{\Sigma_m}
	&= \frac{\sum_i \mu_i\!\left[\dot{\varphi}_i W_i + \frac{\I}{m}\dot{R}_i(\p W/\p R)_i\right]\!\Exp{-\I m\varphi_i}}
	{\sum_i \mu_i W_i\,\Exp{-\I m\varphi_i}}.
\end{align}
An equivalent expression using only real-valued arithmetic is provided in Appendix~\ref{app:Omega}, which also specifies the estimation of the statistical uncertainty $\sigma_\Omega$. This approach trivially generalises to more general windows $W(R,z)$.

\subsection{Moment-of-inertia methods}
\label{sec:MoI}
We define a generalised two-dimensional moment-of-intertia
\begin{align}
	\tens{M} = \int\!\diff^3\!\vec{x}\,\rho(\vec{x},t)\,W(R)\,
		\begin{pmatrix} \hat{x}^2 & \hat{x}\hat{y} \\ \hat{x}\hat{y} & \hat{y}^2		\end{pmatrix}
	\quad\text{with $\hat{x},\hat{y}\equiv\frac{x,y}{R}$}.
\end{align}
The symmetric matrix $\tens{M}$ has orthonormal eigenvectors 
\begin{align}
    \vec{e}_1 = \begin{pmatrix} \cos\psi  \\ \sin\psi \end{pmatrix},\quad
    \vec{e}_2 = \begin{pmatrix} -\sin\psi \\ \phantom{-} \cos\psi \end{pmatrix}
\end{align}
that rotate with the bar as $\dot{\vec{e}}_1=\Omega\vec{e}_2$ and $\dot{\vec{e}}_2=-\Omega\vec{e}_1$ with $\Omega=\dot\psi$. Moreover, with $\lambda_i$ the eigenvalue associated with $\vec{e}_i$, $\tens{M}$ can be written
\begin{align}
    \label{eq:M=sumlee}
    \tens{M} = \sum_{i}\lambda_i \vec{e}_i\otimes\vec{e}_i,
\end{align}
where $\otimes$ denotes the usual outer (or tensor) product. Differentiating with respect to time gives
\begin{align}
    \label{eq:dotM:2D}
    \dot{\tens{M}} = \sum_{i}
    \dot{\lambda}_i \vec{e}_i\otimes\vec{e}_i \quad + \quad
    \Omega(\lambda_1-\lambda_2)(\vec{e}_1\otimes\vec{e}_2 + \vec{e}_2\otimes\vec{e}_1).
\end{align}
Multiplying from left by $\vec{e}_1$ and from right by $\vec{e}_2$ and using their orthonormality, we find
\begin{align}
	\label{eq:Omega:m=2:W:A}
    \Omega &= \frac{\vec{e}_1\cdot\dot{\tens{M}}\cdot\vec{e}_2}{\lambda_1-\lambda_2}.
\end{align}
This relation for $\Omega$ is in fact \emph{identical} to equation~\eqref{eq:Omega:Fourier} for $m=2$, as one can verify by expressing $\lambda_i$ and $\vec{e}_i$ in terms of the matrix elements $\mathsf{M}_{i\!j}$ and exploiting $\cos2\varphi=\hat{x}^2-\hat{y}^2$, $\sin2\varphi=2\hat{x}\hat{y}$. For $W=R^2$, $\tens{M}$ is the moment of inertia of the whole system, when equation~\eqref{eq:Omega:m=2:W:A} agrees with equation~(12) of \citeauthor{WuPfennigerTaam2018}, who did not notice the close relation to the $m=2$ Fourier method.

The relative $m=2$ Fourier amplitude can be expressed in terms of the eigenvalues as  (assuming the order $\lambda_1 \ge \lambda_2$)
\begin{align}
    \label{eq:A2:lambda}
    A_2 \equiv \frac{\Sigma_2}{\Sigma_0}
    = \frac{\left|\int\diff^3\!\vec{x}\rho W\Exp{-2\I\varphi}\right|}{\int\diff^3\!\vec{x}\rho W}
    = \frac{\lambda_1-\lambda_2}{\lambda_1+\lambda_2}.
\end{align}

The concept of rotation of the eigenvectors of a generalised moment of inertia easily extends to three dimensions. To this end, we supplement the matrix $\tens{M}$ with the $z$-coordinate and take the window to be a function of spherical radius $r$:
\begin{align}
	\label{eq:M}
	\tens{M} = \int\!\diff^3\!\vec{x}\,\rho(\vec{x},t)\,W(r)\,
		\uvec{x}\otimes\uvec{x}
	\qquad\text{with $r=|\vec{x}|$ and $\uvec{x}\equiv\vec{x}/r$}
\end{align}
with time derivative
\begin{align}
	&\dot{\tens{M}} = \int\!\diff^3\!\vec{x}\,\rho
		\left[\vec{v}\cdot\uvec{x}\left(\tdiff{W}{r}-2\frac{W}{r}\right) \uvec{x}\otimes\uvec{x}
		+ \frac{W}{r} \big(\vec{v}\otimes\uvec{x} + \uvec{x}\otimes\vec{v}\big)
		\right].
\end{align}
Again, the discrete forms for $\tens{M}$ and $\dot{\tens{M}}$ are easily obtained via the substitutions $\int\diff^3\!\vec{x}\rho \to \sum_i\mu_i$ and $\int\diff^3\!\vec{x}\rho\bar{\vec{v}} \to \sum_i\mu_i\vec{v}_i$. The three orthonormal eigenvectors of $\tens{M}$ form a triad rotating as
\begin{align}
    \label{eq:rot:ei}
    \dot{\vec{e}}_i=\vec{\Omega}\cross\vec{e}_i,
\end{align}
where the vector $\vec{\Omega}$ points along the axis of rotation, while its magnitude is the rotation rate. We again have equation~\eqref{eq:M=sumlee} with time derivative
\begin{align}
    \label{eq:dotM:3D}
    \dot{\tens{M}} = \sum_{i}
    \dot{\lambda}_i \vec{e}_i\otimes\vec{e}_i +
    \sum_{\mathrm{cyclic}}
    \vec{\Omega}\cdot\vec{e}_1(\lambda_2-\lambda_3)(\vec{e}_2\otimes\vec{e}_3 + \vec{e}_3\otimes\vec{e}_2),
\end{align}
when multiplying from left and right by $\vec{e}_i$ and $\vec{e}_{\!j}$, we find
\begin{align}
	\label{eq:Omega:3D}
	&\vec{\Omega} = 
		\frac{\vec{e}_2\cdot\dot{\tens{M}}\cdot\vec{e}_3}{\lambda_2-\lambda_3}
		\vec{e}_1
	+	\frac{\vec{e}_3\cdot\dot{\tens{M}}\cdot\vec{e}_1}{\lambda_3-\lambda_1}
		\vec{e}_2
	+	\frac{\vec{e}_1\cdot\dot{\tens{M}}\cdot\vec{e}_2}{\lambda_1-\lambda_2}
		\vec{e}_3.
\end{align}
The component of $\vec{\Omega}$ in direction $\vec{e}_i$ is only well-defined if the eigenvalues associated with the other two eigenvectors are distinct. Geometrically, this simply states that if a system is axially symmetric, the pattern speed for rotation around the symmetry axis is ill-defined.

One may define a relative $m=2$ Fourier amplitude with respect to each eigenvector as axis. These are given in terms of the eigenvalues via (assuming the order $\lambda_1 \ge \lambda_2 \ge \lambda_3$)
\begin{align}
    \label{eq:A2:lambda:3D}
    A_{2,1} &= \frac{\lambda_2-\lambda_3}{\lambda_2+\lambda_3}, &
    A_{2,2} &= \frac{\lambda_1-\lambda_3}{\lambda_1+\lambda_3}, &
    A_{2,3} &= \frac{\lambda_1-\lambda_2}{\lambda_1+\lambda_2},
\end{align}
when $A_{2,3}$ agrees with the azimuthal Fourier result~\eqref{eq:A2:lambda} for $\uvec{z}=\vec{e}_3$, as expected. The time derivatives of these relative amplitudes can be obtained from the rates of change of the eigenvalues
\begin{align}
    \dot{\lambda}_i= \vec{e}_i\cdot\dot{\tens{M}}\cdot\vec{e}_i
\end{align}
and provide a measure of the non-stationarity of the pattern. 

\begin{figure}
    \includegraphics[width=\columnwidth]{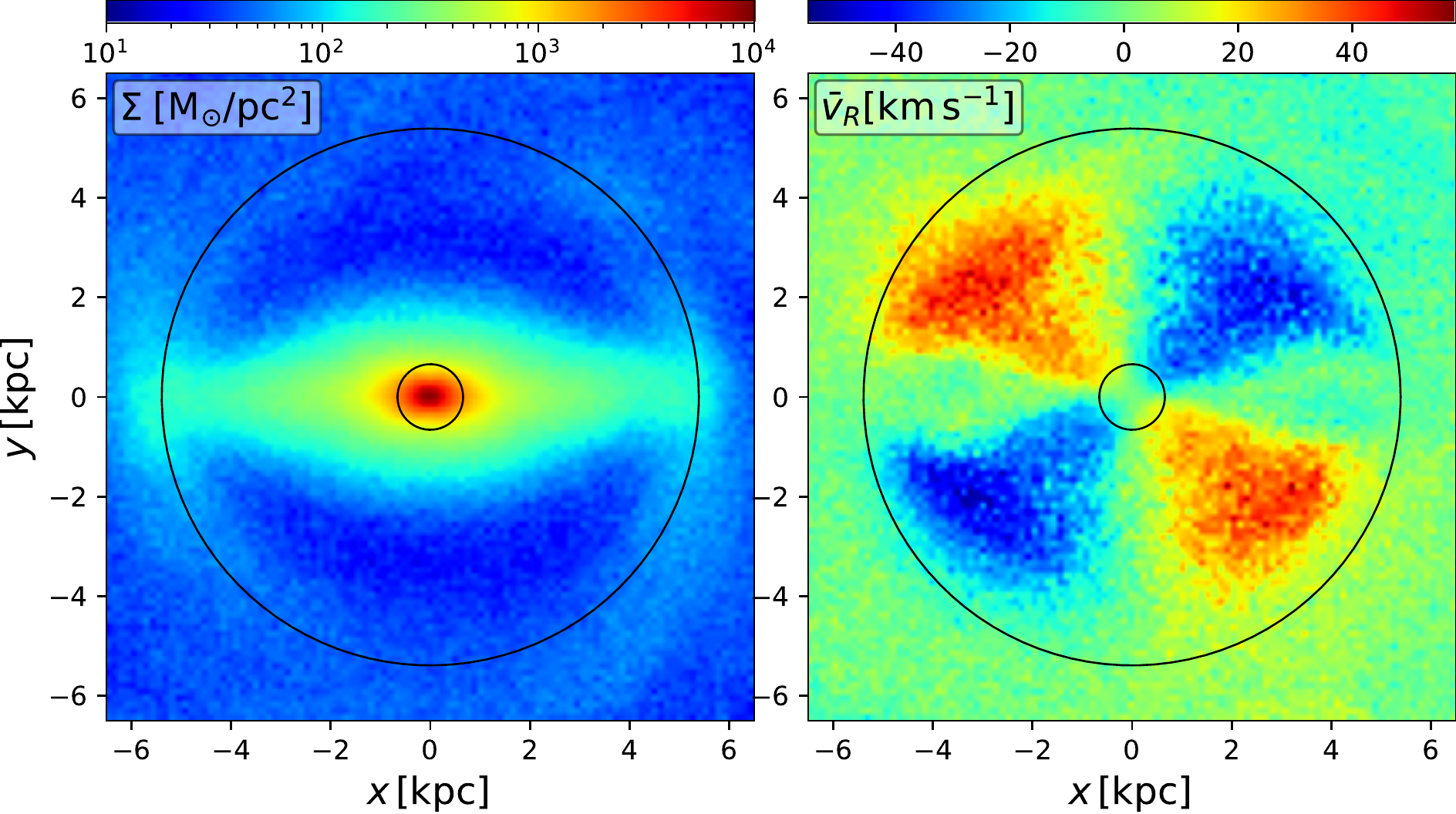}
    \vspace*{-4mm}
    \caption{Surface density (left) and mean radial velocity (right) of our fiducial barred $N$-body model at $t=8\,$Gyr (see Fig.~\ref{fig:maps}). The bar region identified by the method of Appendix~\ref{app:details} is indicated by circles. The radial flux into this region has a $\sin2\varphi$ pattern and must not be neglected when estimating $\Omega$.
    \label{fig:vmean}}
\end{figure}

\begin{figure*}
    \includegraphics[height=4.6cm]{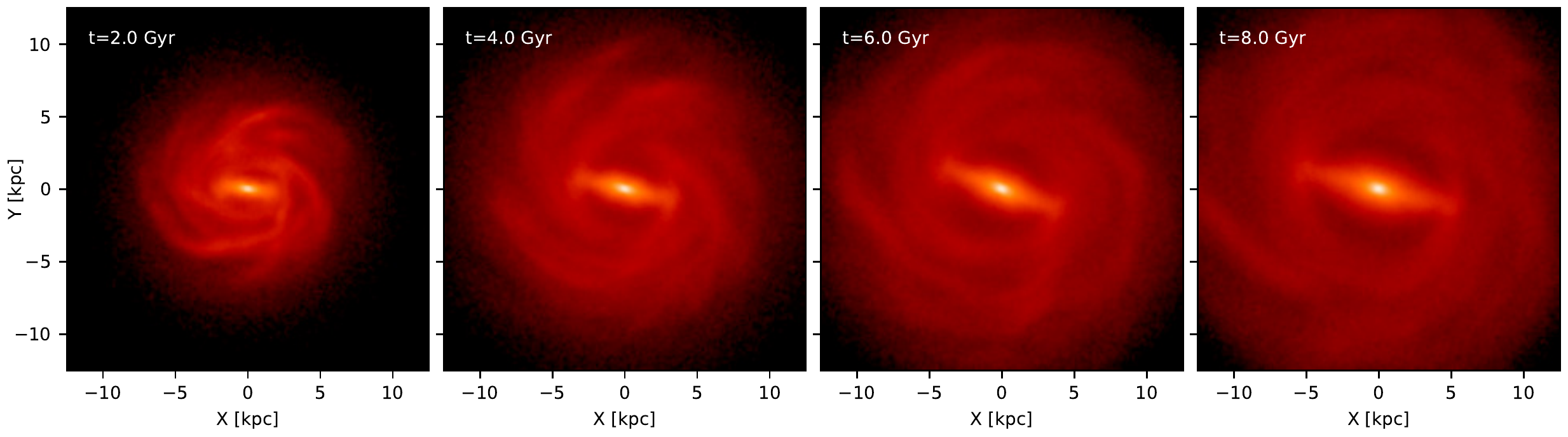}
    \vspace*{-2mm}
    \caption{Snapshots of the stellar surface density for our fiducial growing-disc model. Note the growth of disc and bar as well as the relative strength of spiral arms.}
    \label{fig:maps}
\end{figure*}

\subsection{Particle selection and systematic errors}
\label{sec:bias}
When selecting particles for measuring $\Omega$, for example those in the bar region, one must distinguish the \emph{instantaneous} selection of particles currently in the bar region from the \emph{evolving} selection of particles that are in the bar region at any given time. Since the bar orientation $\psi$ is measured from all particles inside the bar region at the time of measurement, its rate of change $\Omega$ must be measured from the evolving selection to be consistent.

In our method, the bar region is selected via the window function $W(R)$ and the term involving $v_R(\p W/\p R)$ in  equation~\eqref{eq:Omega:Fourier} accounts for the difference between the instantaneous and evolving selections. We demonstrate this by considering a top-hat window, when $W=1$ for $R_0\le R\le R_1$ and zero otherwise, representing an annulus. Then $\p W/\p R=\delta(R-R_0)-\delta(R-R_1)$ and equation~\eqref{eq:Omega:Fourier} gives for $m=2$
\begin{align}
    \label{eq:Omega:Fourier:annulus}
    \Omega =
    \frac{\int\diff \varphi \left(\cos2\varphi\int_{R_0}^{R_1}\diff R\, R\bar{\dot{\varphi}}\Sigma-\tfrac12 \sin2\varphi \big[R \bar{v}_R\Sigma \big]^{R_1}_{R_0}\right)}
    {\int\diff\varphi\cos 2\varphi\int_{R_0}^{R_1} \diff R\,R\Sigma},
\end{align}
where the mean velocity is averaged over $z$ and the coordinate system aligned with the bar. The second term in the numerator accounts for the flux of particles into and out of the window. As we show in Fig.~\ref{fig:vmean}, this flux is generally non-zero, owing to the motion of stars along the bar, and has a $\sin2\varphi$ pattern. Previous authors have \new{either omitted this term \citep{WuPfennigerTaam2018} and thereby implemented the wrong instantaneous selection, or have estimated it only approximately \citep{Frankel2022}, both are prone to systematic errors for $\Omega$.}

For particle systems, this term cannot be easily evaluated (since the chance to find a particle at $R_0$ or $R_1$ is zero). The natural way to solve this problem, is to use a smooth window function (similar to SPH), for example
\begin{align}
    \label{eq:W(R)}
    W(R) = (1-Q)^2(1+2Q)
    \quad\text{with}\quad
    Q = \frac{R^2-\sub{R}{m}^2}{\sub{R}{e}^2-\sub{R}{m}^2},
\end{align}
where $\sub{R}{e}=R_0$ for $R<\sub{R}m$ and $R_1$ for $R>\sub{R}m$, some radius in the middle of the annulus, such as the median.

Similar problems arise whenever selecting particles based on evolving properties (position, velocity, age). For example, measuring $\Omega$ from particles currently on the left side of the galaxy may obtain a biased answer for $\Omega$ on the left size of the galaxy. Such issues are avoided when basing the selection on conserved quantities, such as stellar birth properties (time, location, metallicity), when the instantaneous and evolving selections coincide.

A similar issue occurs if the bar region changes on account of bar evolution. In this case, the window function $W=W(R,t)$ and its time derivative must also be taken into account for a strict  implement of the evolving selection. However, since the pattern speed should not differ between different parts of the bar and because determining the bar region with differentiable edges $R_{0,1}(t)$ is non-trivial, we neglect these terms but do not find significant deviations from $\psi=\int\Omega\diff t$.

\subsection{Measuring the bar pattern speed}
\label{sec:method:bar}
The window for measuring the bar pattern speed should contain most of the bar and not much else. We identify the bar region $[R_0,R_1]$ as a continuous range of radial bins with large $A_2$ and similar $\psi_2$, see Appendix~\ref{app:details} for details. The bar orientation $\psi$ and pattern speed $\Omega$ is measured using the $m=2$ Fourier method from all star and gas particles in the bar region using the window function~\eqref{eq:W(R)} with $\sub{R}m$ taken to be the median radius in the bar region.

Owed to the high velocity dispersion for motion along the bar, the flux term $\rho\bar{v}_R(\partial W/\partial R)\Exp{-\I m\varphi}$ in the numerator of equation~\eqref{eq:Omega:Fourier} contributes non-neglibibly to the  statistical uncertainty $\sigma_\Omega$. In order to reduce this contribution, the window~\eqref{eq:W(R)} is near-maximally smooth. For less smooth window functions (with a larger central part of $W=1$ and steeper $\partial W/\partial R$ at the edges), we find larger $\sigma_\Omega$ and larger deviations to the finite-difference estimate for $\Omega$.

\section[]{\boldmath Testing on $N$-body models}
\label{sec:test}
We test our methods for identifying the bar and measuring its pattern speed on a set of $N$-body models previously used by \cite{Semczuk2022}. We briefly describe these models, before presenting the test results.

\subsection[]{\boldmath The $N$-body models}
Our models are generated using the `growing-disc' technique or \cite{AumerSchoenrich2015}, by which star particles are continuously added to a running $N$-body model. We follow the star formation recipe of \citeauthor{AumerSchoenrich2015} by placing stars on near circular orbits with velocity dispersion of 10\,km\,s$^{-1}$. The total star-formation rate is initially $16.7\;\mathrm{M_{\odot}}/\mathrm{yr}$ and decays exponentially with a decay time scale of 8\,Gyr. The spatial distribution of new stars follows an exponential disc profile with scale length growing from 0.6 to 3\,kpc in 10\,Gyr \citep[following equation 1 of][]{SchoenrichMcmillan2017}, to emulate an inside-out growth. Once the bar is formed, as inferred from the on-fly $m=2$ Fourier analysis, star formation is halted at $0.05<R/R_{\mathrm{CR}}<0.7$, where $R_\mathrm{CR}$ is the co-rotation radius. This is done to mimic gas depletion and subsequent suppression of star formation in the bar and causes the star particles to satisfy the continuity equation (which underlies the methods of Section~\ref{sec:method}) in the bar region.

We implemented in the growing-disc technique with our code \textsc{Griffin} that uses the fast multipole method as force solver \citep{Dehnen2000, Dehnen2014}. The growing disc is embedded in a dark matter whose initial distribution follows a spherical \cite{DehnenMclaughlin2005} profile with scale radius $r_{\mathrm{s}}=31.25$ kpc, smoothly truncated at $10r_{\mathrm{s}}$ and with circular velocity at $r_{\mathrm{s}}$ of to 126.6\,km\,s$^{-1}$. Dark-matter particles are seeded from an ergodic distribution function.

\begin{figure}
    \includegraphics[height=4.6cm]{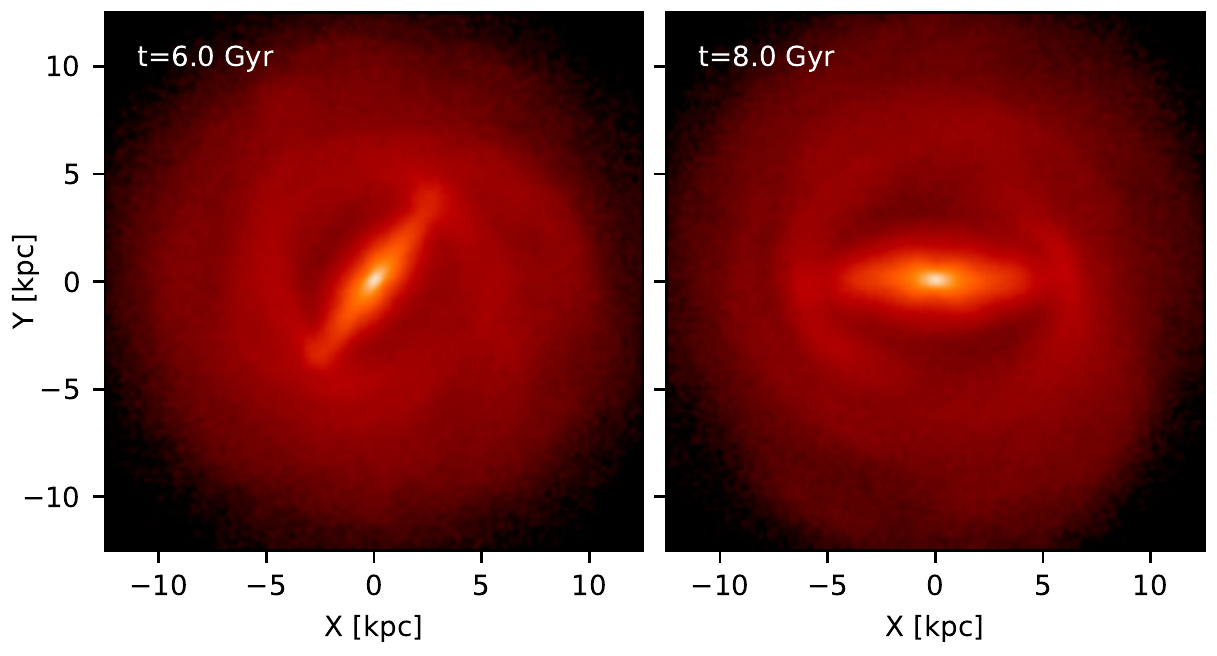}
    \vspace*{-5mm}
    \caption{Snapshots of the stellar surface density from the hot growing-disc model, which is identical to the fiducial model (Fig.~\ref{fig:maps}) at $t\le4\,$Gyr, but has no star formation thereafter.}
    \label{fig:maps_hot}
\end{figure}

We used and tested our method for measuring $\Omega$ on many such growing-disc $N$-body models with different values for the various parameters of dark halo and star formation. However, here we present only two illustrative typical models, which we call `fiducial' and `hot'. In the fiducial model, star formation takes place throughout the whole evolution, while for the hot model it is turned off after 4\,Gyr. Stellar particles have a mass of $1.2\times10^4\;\mathrm{M_{\odot}}$ and softening length $50\,$pc, while halo particles have a mass of $2.7\times10^5\;\mathrm{M_{\odot}}$ and softening length $200\,$pc. The dark matter halo has $4.2\times10^6$ particles, while at $8\,$Gyr the fiducial and hot models have $2.4\times10^6$ and $1.9\times10^6$ star particles, respectively.

Fig.~\ref{fig:maps} shows snapshots of surface density for the fiducial model, which forms a bar within the first Gyr, that then grows, strengthens and slows down -- the typical evolution of bars surrounded by dark matter halos (see introduction). For comparison, Fig.~\ref{fig:maps_hot} shows the hot model at the same last two times. The main difference is the absence of transient spiral structure, which in the fiducial simulation is present at all times on account of the continuous addition of dynamically cold (low Toomre $Q$) material.

\begin{figure}
    \includegraphics[width=\columnwidth]{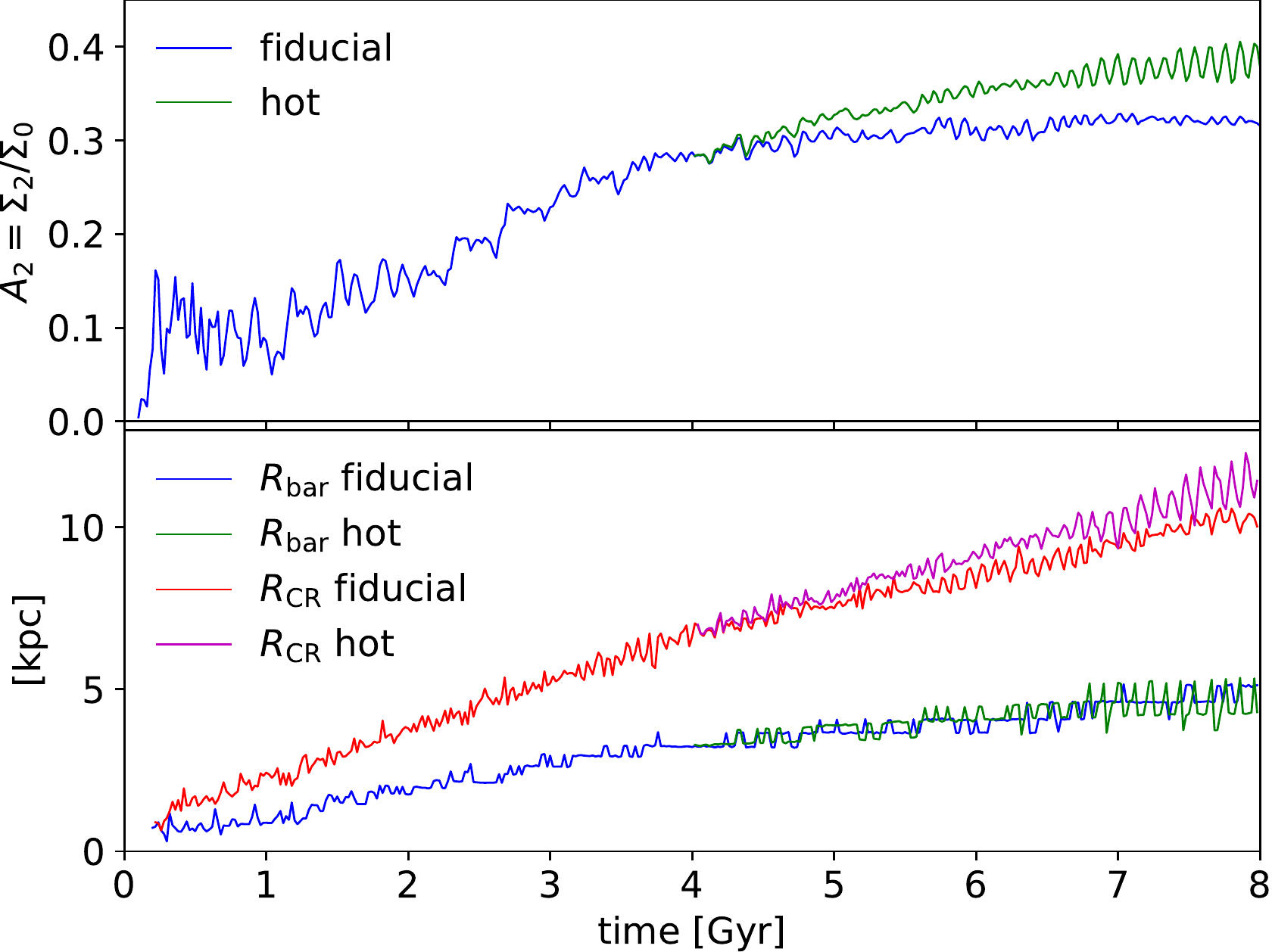}
    \vspace*{-5mm}
    \caption{Time evolution of the relative $m=2$ Fourier amplitude measured from all particles inside 5\,kpc (top) and of the bar length and the co-rotation radius for the fiducial and hot $N$-body models.}
    \label{fig:evolution}
\end{figure}

Figure~\ref{fig:evolution} shows the time evolution of the relative $m=2$ Fourier amplitude $A_2=\Sigma_{m=2}/\Sigma_{m=0}$ as well as bar size $\sub{R}{bar}$ and co-rotation radius $\sub{R}{CR}$ of both models. After $\sim5\,$Gyr, the growth of bar strength and size (as measured by $A_2$ and $\sub{R}{bar}$) is significantly reduced, while the co-rotation radius continues to rise roughly linearly, corresponding to a decelerating bar (since $\Omega\sim1/\sub{R}{CR}$). The hot differs from the fiducial model in that its bar slows down slightly more and also grows stronger and longer, in agreement with the observation that bars in earlier type galaxies tend to be stronger and longer \citep[but see \citealt{DiazGarcia2016}]{Elmegreen1985, Erwin2005, Gadotti2011}.

\begin{figure}
    \includegraphics[width=\columnwidth]{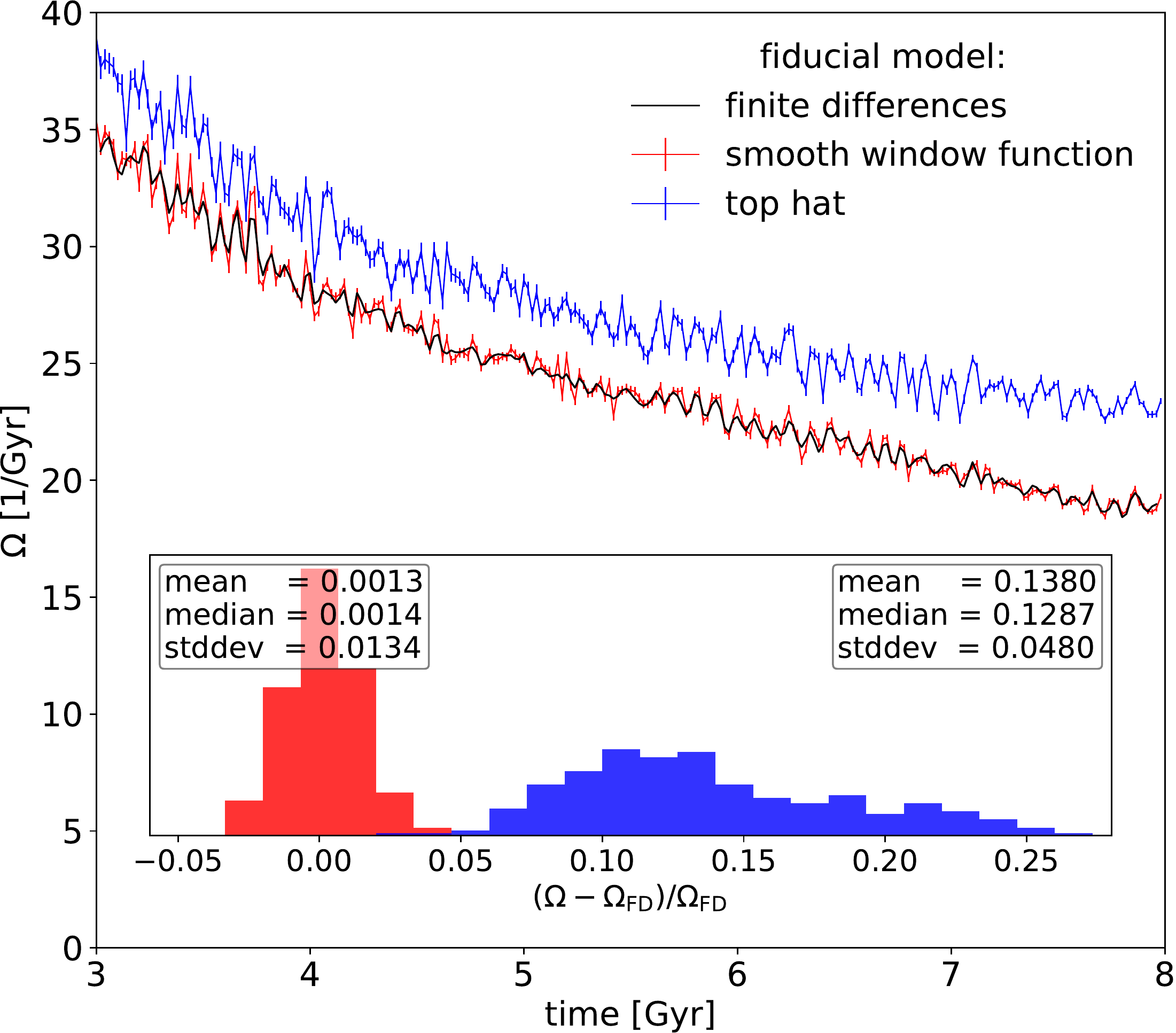}
    \vspace*{-4mm}
    \caption{Bar pattern speed measured for the fiducial simulation with the $m=2$ Fourier method using a smooth (red) or top-hat (blue) window function, or as finite difference (equation~\ref{eq:Om:FD}) with $\Delta t=40\,$Myr from $\psi$ obtained via the smooth window (using the top-hat window for this purpose makes no appreciable difference). Error bars are estimated from particle noise. The inlet shows the distributions of relative differences of the instantaneously measured pattern speeds to the finite difference estimate with legends reporting their mean, median and standard deviation.}
    \label{fig:test0}
\end{figure}

\subsection{Testing the instantaneous Fourier method}
\label{sec:test:fourier}
We first test our method on the fiducial model in the time window 3-8\,Gyr, when the bar is already well established and slowing down. In Fig.~\ref{fig:test0}, we compare the measurements of $\Omega$ with the commonly used time-centred finite-difference estimate~\eqref{eq:Om:FD} with $\Delta t=40\,$Myr (black). Note that in view of equation~\eqref{eq:psi=intOmega} this is just the sliding average
\begin{align}
    \Omega_{\mathrm{FD}}(t) = 
    \frac1{\Delta t} \int_{t-\Delta t/2}^{t+\Delta t/2} \Omega(t') \diff t'.
\end{align}
All three measurements of the pattern speed show a general slowdown and fluctuations with amplitude $\sim5\%$ on time scales of $\sim100\,$Myr.

The naive approach with a top-hat window function ($W=1$ for $R_0<R<R_1$ and $W=0$ otherwise) and ignoring the boundary terms (blue, in Fig.~\ref{fig:test0}) is obviously wrong: it estimates the pattern speed 6-25\% too high with an average bias of $14\%$.

Our estimate for $\Omega$ (red) fluctuates around $\sub{\Omega}{FD}$ (on top of its variations), but shows no significant systematic bias -- the mean deviation of $0.13\%$ is insignificant. The amplitude of the fluctuations of $1.34\%$ is about twice the mean relative uncertainty $\langle\sigma_\Omega/\sub{\Omega}{FD}\rangle=0.75\%$ due to particle shot noise ($\sigma_\Omega$ is shown as error-bars in the top panel of Fig.~\ref{fig:test0}). This suggests that the fluctuations of the instantaneously measured $\Omega$ on time scales $<\Delta t=40\,$Myr are only partly due to shot noise, but mostly reflect true variations of $\Omega$ on these time scales. On time scales much shorter than the shot-noise correlation time, the instantaneously measured $\Omega$ and $\Omega_{\mathrm{FD}}$ do indeed agree (if the bar region is kept fixed), as we verify using a simulation with output interval $10^4$ times shorter (not plotted).

If we decrease the number of particles used to estimate $\Omega$ by a factor six, the standard deviation of the relative deviations to $\sub{\Omega}{FD}$ is $1.94\%$, not quite twice the value reported in Fig.~\ref{fig:test0}, and the measured statistical uncertainty $\sigma_\Omega$ rises to $1.75\%$ (from $0.72\%$) -- close to the expected rise by $\sqrt{6}\approx2.45$.

\begin{figure}
    \includegraphics[width=\columnwidth]{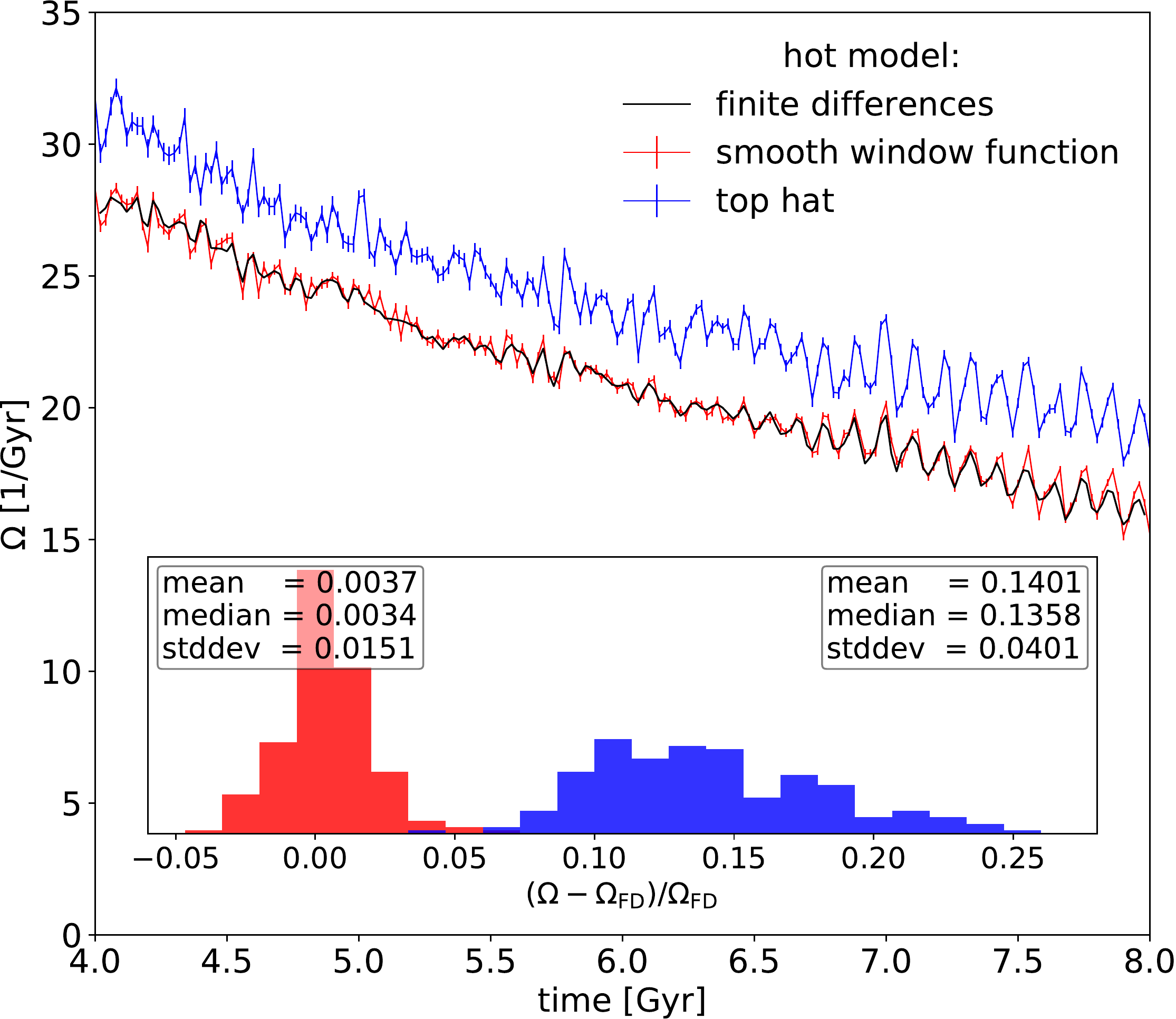}
    \vspace*{-4mm}
    \caption{As Fig.~\ref{fig:test0} but for the hot model.}
    \label{fig:hot_test}
\end{figure}

In Fig.~\ref{fig:hot_test}, we show pattern speeds measured for the hot model (for $t>4\,$ Gyr, when it differs from the fiducial simulation). Our method is unbiased, while the naive approach incurs a $\sim14\%$ bias, corroborating our findings for the fiducial model. We also clearly see that the pattern-speed oscillations at $t=7$-8\,Gyr have substantially larger amplitude for $\Omega$ measured instantaneously than its sliding average, as provided by the finite difference.

\begin{figure}
    \includegraphics[width=\columnwidth]{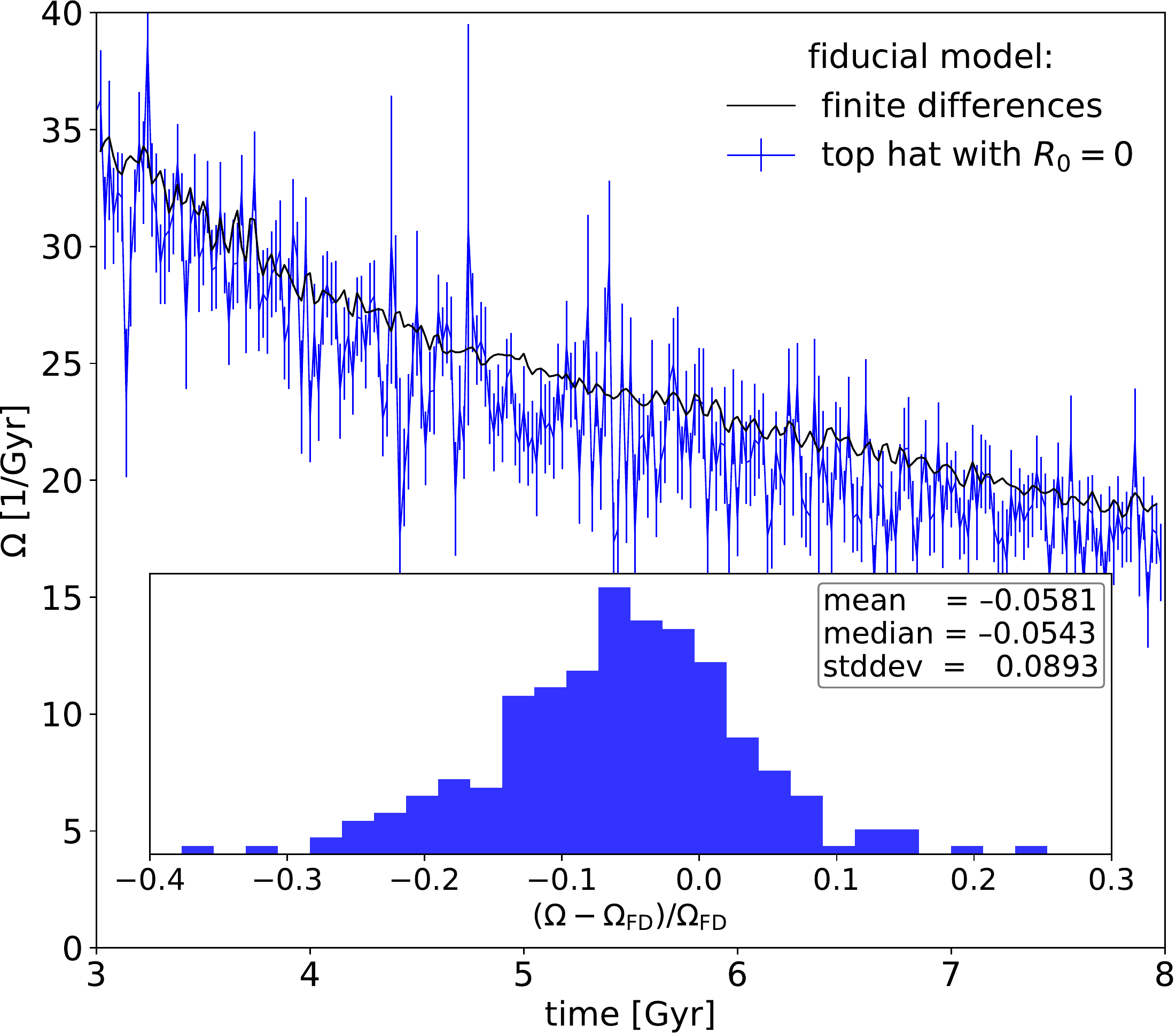}
    \vspace*{-4mm}
    \caption{As Fig.~\ref{fig:test0} for the top-hat window function with inner edge $R_0=0$.}
    \label{fig:test0:Rmin=0}
\end{figure}

In Fig.~\ref{fig:test0:Rmin=0}, we use the naive method with a bar region without central hole, i.e.\ $R_0=0$. This has the advantage that the bar region no longer has an inner boundary such that the neglected flux is limited to that through the outer boundary. We find that the bias of the method has changed sign and is smaller compared to the situation with inner boundary (in Fig.~\ref{fig:test0}). However, we also see that the method has become very noisy. Upon inspection of equation~\eqref{eq:Omega:Fourier:i} this is not surprising, since $\dot{\varphi}=v_\varphi/R$ can become arbitrarily large for particles near $R=0$. We conclude from this exercise, that for the purpose of calculating $\Omega$, the bar region must exclude the origin and therefore necessarily have an inner boundary, though our smooth-window approach suffers much less from this problem, as it weighs particles at $R\sim R_0$ only very little.

\section{Discussion}
\label{sec:discussion}
Rotating galactic bars are always dominated by their azimuthal $m=2$ Fourier mode. Consequently, their orientation $\psi$ is well estimated by the phase of the $m=2$ Fourier transform measured in the bar region. The bar pattern speed, therefore, is naturally defined as the time derivative of that phase.

We show in Section~\ref{sec:bias} and demonstrate in Section~\ref{sec:test:fourier} that in order to compute $\Omega=\diff\psi/\diff t$ without bias, one must account for the net particle flux into the bar region. Previous implementations of this approach \new{ either overlooked this flux term \citep{WuPfennigerTaam2018}, resulting in large systematic errors, or implemented it only approximately \citep{Frankel2022}, when the measured $\psi$ and $\Omega$ are inconsistent, i.e.\ do not in general satisfy $\psi=\int\Omega\diff t$. This flux term, which} never vanishes, is most naturally determined \new{in a consistent way by} weighing the simulation particles \new{with} a function $W(R)$ that smoothly drops to zero outside the bar region.

When attempting to use the Tremaine-Weinberg method to separately measure pattern speeds for the inner and outer parts of an external galaxy, a similar flux term, which cannot be measured, occurs at their boundary. Such applications are therefore erroneous.

In Section~\ref{sec:MoI}, we show that the Fourier method is equivalent, modulo a radial weight function, to measuring $\Omega$ as the rate at which the eigenvectors of the planar moment of inertia rotate. We also show that this method can be generalised to obtain the \emph{vector} $\vec{\Omega}$, the rate \emph{and} axis of rotation, as that by which the triad of the principal axes of a 3D moment-of-inertia-like tensor rotates. We have not tested this generalisation on $N$-body models of galactic bars (for which the direction of rotation is unambiguous), but suggest it for measuring the tumbling rate and axis of spheroidal components, such as elliptical galaxies and dark-matter haloes.

While defining the bar orientation $\psi$ as phase of the $m=2$ azimuthal Fourier mode appears natural and works well, alternative ideas are worth considering. One is to trace the phase of the azimuthal maximum, i.e.\ the ridge of the bar, which is more like what humans do when eye-balling the bar orientation. To find this maximum some azimuthal smoothing is necessary using a smoothing kernel $w(\Delta\varphi)$, when only particles near the maximum contribute to the estimate. Such an approach is therefore likely to be more noisy, unless the azimuthal smoothing is maximally wide, like $w=\frac12[1+\cos2\Delta\varphi]$, when it reverts to the $m=2$ Fourier method.

Can one determine $\Omega=\dot\psi$ without any concept of the bar orientation $\psi$? This seems impossible, but \citeauthor{WuPfennigerTaam2018} also proposed a method which does, by estimating $\Omega$ as the value for which the Jacobi integral $J=E-\Omega L_z$ is conserved (giving $\Omega=\dot{E}/\dot{L}_{z}$ with least-squares solution $\Omega=\sum_i\dot{E}_i\dot{L}_{z,i}/\sum_i\dot{L}_{z,i}^2$). Unfortunately, this clever approach has two serious problems. First, $J$ is conserved only if the gravitational potential is stationary (not just the pattern as required by the Tremaine-Weinberg method) and $\Omega$ is constant in time. Neither of these conditions is likely satisfied for realistic simulations and systematic errors are unavoidable. Second, this method requires knowledge of the time derivative of the gravitational potential for all particles, which is not usually computed by $N$-body force solvers.

All these considerations strongly favour the $m=2$ Fourier method, including its formulation as moment-of-inertia method, over other contemporary approaches for measuring bar pattern speeds.

The task of determining $\Omega$ for spiral structure is harder than for bars, since spirals are weaker, evolve faster, and can contain patterns rotating at different rates. However, our method should in theory be capable to measure their pattern speeds, one for each azimuthal wavenumber $m$ and (sufficiently resolved) radial range, as well as the time derivatives of the wave amplitudes. While we have not attempted or tested this so far, this is a promising idea warranting further investigation.

\section{Conclusions}
\label{sec:conclude}
We provide the first unbiased method for measuring the bar pattern speed $\Omega$ from single simulation snapshots. This is valuable because time intervals between snapshots are typically too long for determining $\Omega$ by following the bar rotation.

\section*{Acknowledgements}
We thank Lia Athanssoula, and Hossam Aly for many stimulating conversations, the reviewer James Binney for a valuable report, and Daniel Pfenniger \new{and Hans-Walter Rix} for informative discussion\new{s}. MS acknowledges support from STFC grant ST/S000453/1. RS thanks the Royal Society for generous support via a University Research Fellowship. Computations were performed using the DiRAC Data Intensive service at Leicester, operated by the University of Leicester IT Services, which forms part of the STFC DiRAC HPC Facility (www.dirac.ac.uk). The equipment was funded by BEIS capital funding via STFC capital grants ST/K000373/1 and ST/R002363/1 and STFC DiRAC Operations grant ST/R001014/1. DiRAC is part of the National e-Infrastructure.

\section*{Data Availability}
Computer code in \texttt{python} for finding the bar region and estimating bar pattern speed from $N$-body data is publicly available at \url{https://github.com/WalterDehnen/patternSpeed}. The simulation data can be shared on reasonable request.

\bibliographystyle{mnras} \bibliography{main}

\begin{thebibliography}{}
\makeatletter
\relax
\def\mn@urlcharsother{\let\do\@makeother \do\$\do\&\do\#\do\^\do\_\do\%\do\~}
\def\mn@doi{\begingroup\mn@urlcharsother \@ifnextchar [ {\mn@doi@}
  {\mn@doi@[]}}
\def\mn@doi@[#1]#2{\def\@tempa{#1}\ifx\@tempa\@empty \href
  {http://dx.doi.org/#2} {doi:#2}\else \href {http://dx.doi.org/#2} {#1}\fi
  \endgroup}
\def\mn@eprint#1#2{\mn@eprint@#1:#2::\@nil}
\def\mn@eprint@arXiv#1{\href {http://arxiv.org/abs/#1} {{\tt arXiv:#1}}}
\def\mn@eprint@dblp#1{\href {http://dblp.uni-trier.de/rec/bibtex/#1.xml}
  {dblp:#1}}
\def\mn@eprint@#1:#2:#3:#4\@nil{\def\@tempa {#1}\def\@tempb {#2}\def\@tempc
  {#3}\ifx \@tempc \@empty \let \@tempc \@tempb \let \@tempb \@tempa \fi \ifx
  \@tempb \@empty \def\@tempb {arXiv}\fi \@ifundefined
  {mn@eprint@\@tempb}{\@tempb:\@tempc}{\expandafter \expandafter \csname
  mn@eprint@\@tempb\endcsname \expandafter{\@tempc}}}

\bibitem[\protect\citeauthoryear{{Aguerri} et~al.,}{{Aguerri}
  et~al.}{2015}]{Aguerri2015}
{Aguerri} J.~A.~L.,  et~al., 2015, \mn@doi [\aap]
  {10.1051/0004-6361/201423383}, \href
  {https://ui.adsabs.harvard.edu/abs/2015A&A...576A.102A} {576, A102}

\bibitem[\protect\citeauthoryear{{Algorry} et~al.,}{{Algorry}
  et~al.}{2017}]{Algorry2017}
{Algorry} D.~G.,  et~al., 2017, \mn@doi [\mnras] {10.1093/mnras/stx1008}, \href
  {https://ui.adsabs.harvard.edu/abs/2017MNRAS.469.1054A} {469, 1054}

\bibitem[\protect\citeauthoryear{{Athanassoula}}{{Athanassoula}}{2003}]{Athanassoula2003}
{Athanassoula} E.,  2003, \mn@doi [\mnras] {10.1046/j.1365-8711.2003.06473.x},
  \href {https://ui.adsabs.harvard.edu/abs/2003MNRAS.341.1179A} {341, 1179}

\bibitem[\protect\citeauthoryear{{Aumer} \& {Sch{\"o}nrich}}{{Aumer} \&
  {Sch{\"o}nrich}}{2015}]{AumerSchoenrich2015}
{Aumer} M.,  {Sch{\"o}nrich} R.,  2015, \mn@doi [\mnras]
  {10.1093/mnras/stv2252}, \href
  {https://ui.adsabs.harvard.edu/abs/2015MNRAS.454.3166A} {454, 3166}

\bibitem[\protect\citeauthoryear{{Cheung} et~al.,}{{Cheung}
  et~al.}{2013}]{Cheung2013}
{Cheung} E.,  et~al., 2013, \mn@doi [\apj] {10.1088/0004-637X/779/2/162}, \href
  {https://ui.adsabs.harvard.edu/abs/2013ApJ...779..162C} {779, 162}

\bibitem[\protect\citeauthoryear{{Chiba}, {Friske}  \& {Sch{\"o}nrich}}{{Chiba}
  et~al.}{2021}]{Chiba2021}
{Chiba} R.,  {Friske} J. K.~S.,   {Sch{\"o}nrich} R.,  2021, \mn@doi [\mnras]
  {10.1093/mnras/staa3585}, \href
  {https://ui.adsabs.harvard.edu/abs/2021MNRAS.500.4710C} {500, 4710}

\bibitem[\protect\citeauthoryear{{Corsini}}{{Corsini}}{2011}]{Corsini2011}
{Corsini} E.~M.,  2011, Memorie della Societa Astronomica Italiana Supplementi,
  \href {https://ui.adsabs.harvard.edu/abs/2011MSAIS..18...23C} {18, 23}

\bibitem[\protect\citeauthoryear{{Debattista} \& {Sellwood}}{{Debattista} \&
  {Sellwood}}{1998}]{DebattistaSellwood1998}
{Debattista} V.~P.,  {Sellwood} J.~A.,  1998, \mn@doi [\apjl] {10.1086/311118},
  \href {https://ui.adsabs.harvard.edu/abs/1998ApJ...493L...5D} {493, L5}

\bibitem[\protect\citeauthoryear{{Dehnen}}{{Dehnen}}{2000}]{Dehnen2000}
{Dehnen} W.,  2000, \mn@doi [\apjl] {10.1086/312724}, \href
  {https://ui.adsabs.harvard.edu/abs/2000ApJ...536L..39D} {536, L39}

\bibitem[\protect\citeauthoryear{{Dehnen}}{{Dehnen}}{2014}]{Dehnen2014}
{Dehnen} W.,  2014, \mn@doi [Computational Astrophysics and Cosmology]
  {10.1186/s40668-014-0001-7}, \href
  {https://ui.adsabs.harvard.edu/abs/2014ComAC...1....1D} {1, 1}

\bibitem[\protect\citeauthoryear{{Dehnen} \& {McLaughlin}}{{Dehnen} \&
  {McLaughlin}}{2005}]{DehnenMclaughlin2005}
{Dehnen} W.,  {McLaughlin} D.~E.,  2005, \mn@doi [\mnras]
  {10.1111/j.1365-2966.2005.09510.x}, \href
  {https://ui.adsabs.harvard.edu/abs/2005MNRAS.363.1057D} {363, 1057}

\bibitem[\protect\citeauthoryear{{D{\'\i}az-Garc{\'\i}a}, {Salo}, {Laurikainen}
   \& {Herrera-Endoqui}}{{D{\'\i}az-Garc{\'\i}a} et~al.}{2016}]{DiazGarcia2016}
{D{\'\i}az-Garc{\'\i}a} S.,  {Salo} H.,  {Laurikainen} E.,   {Herrera-Endoqui}
  M.,  2016, \mn@doi [\aap] {10.1051/0004-6361/201526161}, \href
  {https://ui.adsabs.harvard.edu/abs/2016A&A...587A.160D} {587, A160}

\bibitem[\protect\citeauthoryear{{Elmegreen} \& {Elmegreen}}{{Elmegreen} \&
  {Elmegreen}}{1985}]{Elmegreen1985}
{Elmegreen} B.~G.,  {Elmegreen} D.~M.,  1985, \mn@doi [\apj] {10.1086/162810},
  \href {https://ui.adsabs.harvard.edu/abs/1985ApJ...288..438E} {288, 438}

\bibitem[\protect\citeauthoryear{{Erwin}}{{Erwin}}{2005}]{Erwin2005}
{Erwin} P.,  2005, \mn@doi [\mnras] {10.1111/j.1365-2966.2005.09560.x}, \href
  {https://ui.adsabs.harvard.edu/abs/2005MNRAS.364..283E} {364, 283}

\bibitem[\protect\citeauthoryear{{Erwin}}{{Erwin}}{2018}]{Erwin2018}
{Erwin} P.,  2018, \mn@doi [\mnras] {10.1093/mnras/stx3117}, \href
  {https://ui.adsabs.harvard.edu/abs/2018MNRAS.474.5372E} {474, 5372}

\bibitem[\protect\citeauthoryear{{Eskridge} et~al.,}{{Eskridge}
  et~al.}{2000}]{EskridgeEtal2000}
{Eskridge} P.~B.,  et~al., 2000, \mn@doi [\aj] {10.1086/301203}, \href
  {https://ui.adsabs.harvard.edu/abs/2000AJ....119..536E} {119, 536}

\bibitem[\protect\citeauthoryear{{Fragkoudi}, {Grand}, {Pakmor}, {Springel},
  {White}, {Marinacci}, {Gomez}  \& {Navarro}}{{Fragkoudi}
  et~al.}{2021}]{Fragkoudi2021}
{Fragkoudi} F.,  {Grand} R.~J.~J.,  {Pakmor} R.,  {Springel} V.,  {White}
  S.~D.~M.,  {Marinacci} F.,  {Gomez} F.~A.,   {Navarro} J.~F.,  2021, \mn@doi
  [\aap] {10.1051/0004-6361/202140320}, \href
  {https://ui.adsabs.harvard.edu/abs/2021A&A...650L..16F} {650, L16}

\bibitem[\protect\citeauthoryear{{Frankel} et~al.,}{{Frankel}
  et~al.}{2022}]{Frankel2022}
{Frankel} N.,  et~al., 2022, arXiv e-prints, \href
  {https://ui.adsabs.harvard.edu/abs/2022arXiv220108406F} {p. arXiv:2201.08406}

\bibitem[\protect\citeauthoryear{{Gadotti}}{{Gadotti}}{2011}]{Gadotti2011}
{Gadotti} D.~A.,  2011, \mn@doi [\mnras] {10.1111/j.1365-2966.2011.18945.x},
  \href {https://ui.adsabs.harvard.edu/abs/2011MNRAS.415.3308G} {415, 3308}

\bibitem[\protect\citeauthoryear{{Guo}, {Mao}, {Athanassoula}, {Li}, {Ge},
  {Long}, {Merrifield}  \& {Masters}}{{Guo} et~al.}{2019}]{Guo2019}
{Guo} R.,  {Mao} S.,  {Athanassoula} E.,  {Li} H.,  {Ge} J.,  {Long} R.~J.,
  {Merrifield} M.,   {Masters} K.,  2019, \mn@doi [\mnras]
  {10.1093/mnras/sty2715}, \href
  {https://ui.adsabs.harvard.edu/abs/2019MNRAS.482.1733G} {482, 1733}

\bibitem[\protect\citeauthoryear{{Hilmi} et~al.,}{{Hilmi}
  et~al.}{2020}]{Hilmi2020}
{Hilmi} T.,  et~al., 2020, \mn@doi [\mnras] {10.1093/mnras/staa1934}, \href
  {https://ui.adsabs.harvard.edu/abs/2020MNRAS.497..933H} {497, 933}

\bibitem[\protect\citeauthoryear{{Little} \& {Carlberg}}{{Little} \&
  {Carlberg}}{1991}]{LittleCarlberg1991}
{Little} B.,  {Carlberg} R.~G.,  1991, \mn@doi [\mnras]
  {10.1093/mnras/250.1.161}, \href
  {https://ui.adsabs.harvard.edu/abs/1991MNRAS.250..161L} {250, 161}

\bibitem[\protect\citeauthoryear{{Masters} et~al.,}{{Masters}
  et~al.}{2011}]{Masters2011}
{Masters} K.~L.,  et~al., 2011, \mn@doi [\mnras]
  {10.1111/j.1365-2966.2010.17834.x}, \href
  {https://ui.adsabs.harvard.edu/abs/2011MNRAS.411.2026M} {411, 2026}

\bibitem[\protect\citeauthoryear{{Men{\'e}ndez-Delmestre}, {Sheth},
  {Schinnerer}, {Jarrett}  \& {Scoville}}{{Men{\'e}ndez-Delmestre}
  et~al.}{2007}]{MenendezDelmestreEtAl2007}
{Men{\'e}ndez-Delmestre} K.,  {Sheth} K.,  {Schinnerer} E.,  {Jarrett} T.~H.,
  {Scoville} N.~Z.,  2007, \mn@doi [\apj] {10.1086/511025}, \href
  {https://ui.adsabs.harvard.edu/abs/2007ApJ...657..790M} {657, 790}

\bibitem[\protect\citeauthoryear{{Peschken} \& {{\L}okas}}{{Peschken} \&
  {{\L}okas}}{2019}]{Peschken2019}
{Peschken} N.,  {{\L}okas} E.~L.,  2019, \mn@doi [\mnras]
  {10.1093/mnras/sty3277}, \href
  {https://ui.adsabs.harvard.edu/abs/2019MNRAS.483.2721P} {483, 2721}

\bibitem[\protect\citeauthoryear{{Power}, {Navarro}, {Jenkins}, {Frenk},
  {White}, {Springel}, {Stadel}  \& {Quinn}}{{Power} et~al.}{2003}]{Power2003}
{Power} C.,  {Navarro} J.~F.,  {Jenkins} A.,  {Frenk} C.~S.,  {White} S.~D.~M.,
   {Springel} V.,  {Stadel} J.,   {Quinn} T.,  2003, \mn@doi [\mnras]
  {10.1046/j.1365-8711.2003.05925.x}, \href
  {https://ui.adsabs.harvard.edu/abs/2003MNRAS.338...14P} {338, 14}

\bibitem[\protect\citeauthoryear{{Sch{\"o}nrich} \& {McMillan}}{{Sch{\"o}nrich}
  \& {McMillan}}{2017}]{SchoenrichMcmillan2017}
{Sch{\"o}nrich} R.,  {McMillan} P.~J.,  2017, \mn@doi [\mnras]
  {10.1093/mnras/stx093}, \href
  {https://ui.adsabs.harvard.edu/abs/2017MNRAS.467.1154S} {467, 1154}

\bibitem[\protect\citeauthoryear{{Sellwood}}{{Sellwood}}{1980}]{Sellwood1980}
{Sellwood} J.~A.,  1980, \aap, \href
  {https://ui.adsabs.harvard.edu/abs/1980A&A....89..296S} {89, 296}

\bibitem[\protect\citeauthoryear{{Sellwood} \& {Athanassoula}}{{Sellwood} \&
  {Athanassoula}}{1986}]{Sellwood1986}
{Sellwood} J.~A.,  {Athanassoula} E.,  1986, \mn@doi [\mnras]
  {10.1093/mnras/221.2.195}, \href
  {https://ui.adsabs.harvard.edu/abs/1986MNRAS.221..195S} {221, 195}

\bibitem[\protect\citeauthoryear{{Semczuk}, {Dehnen}, {Sch{\"o}nrich}  \&
  {Athanassoula}}{{Semczuk} et~al.}{2022}]{Semczuk2022}
{Semczuk} M.,  {Dehnen} W.,  {Sch{\"o}nrich} R.,   {Athanassoula} E.,  2022,
  \mn@doi [\mnras] {10.1093/mnras/stab3294}, \href
  {https://ui.adsabs.harvard.edu/abs/2022MNRAS.509.4532S} {509, 4532}

\bibitem[\protect\citeauthoryear{{Sheth} et~al.,}{{Sheth}
  et~al.}{2008}]{ShethEtAl2008}
{Sheth} K.,  et~al., 2008, \mn@doi [\apj] {10.1086/524980}, \href
  {https://ui.adsabs.harvard.edu/abs/2008ApJ...675.1141S} {675, 1141}

\bibitem[\protect\citeauthoryear{{Tremaine} \& {Weinberg}}{{Tremaine} \&
  {Weinberg}}{1984}]{TremaineWeinberg1984}
{Tremaine} S.,  {Weinberg} M.~D.,  1984, \mn@doi [\apjl] {10.1086/184292},
  \href {https://ui.adsabs.harvard.edu/abs/1984ApJ...282L...5T} {282, L5}

\bibitem[\protect\citeauthoryear{{Weinberg}}{{Weinberg}}{1985}]{Weinberg1985}
{Weinberg} M.~D.,  1985, \mn@doi [\mnras] {10.1093/mnras/213.3.451}, \href
  {https://ui.adsabs.harvard.edu/abs/1985MNRAS.213..451W} {213, 451}

\bibitem[\protect\citeauthoryear{{Wu}, {Pfenniger}  \& {Taam}}{{Wu}
  et~al.}{2016}]{Wu2016}
{Wu} Y.-T.,  {Pfenniger} D.,   {Taam} R.~E.,  2016, \mn@doi [\apj]
  {10.3847/0004-637X/830/2/111}, \href
  {https://ui.adsabs.harvard.edu/abs/2016ApJ...830..111W} {830, 111}

\bibitem[\protect\citeauthoryear{{Wu}, {Pfenniger}  \& {Taam}}{{Wu}
  et~al.}{2018}]{WuPfennigerTaam2018}
{Wu} Y.-T.,  {Pfenniger} D.,   {Taam} R.~E.,  2018, \mn@doi [\apj]
  {10.3847/1538-4357/aac5e8}, \href
  {https://ui.adsabs.harvard.edu/abs/2018ApJ...860..152W} {860, 152}

\makeatother
\end{thebibliography}

\appendix

\section{Real-valued Fourier method}
\label{app:Omega}
We begin by noting that the azimuthal harmonics are best recursively computed from $\cos\varphi=x/R$, $\sin\varphi=y/R$,
\begin{subequations}
\begin{align}
	\cos[m+1]\varphi &=\cos m\varphi\, \cos\varphi - \sin m\varphi\,\sin\varphi,
	&\text{and} \\
	\sin[m+1]\varphi &=\cos m\varphi\, \sin\varphi + \sin m\varphi\,\cos\varphi
\end{align}
\end{subequations}
(in particular $\cos2\varphi=(x^2-y^2)/R^2$ and $\sin2\varphi=2xy/R^2$), which is computationally much faster than calls to trigonometric functions. Given the sums
\begin{align}
	\label{eq:Cm,Sm}
	C_m &\equiv \sum_i\mu_i W(\vec{x}_i) \cos m\varphi_i,&
	S_m &\equiv \sum_i\mu_i W(\vec{x}_i) \sin m\varphi_i,
\end{align}
with $\partial W/\partial\varphi=0$ and their time derivatives
\begin{subequations}
	\label{eqs:dt:Cm,Sm}
\begin{align}
	\dot{C}_m &= \sum_i \mu_i \left[\vec{v}_i\cdot\vec{\nabla}W(\vec{x}_i) \cos m\varphi_i - m\dot{\varphi}_i W(\vec{x}_i) \sin m\varphi_i \right], \\
	\dot{S}_m &= \sum_i \mu_i \left[\vec{v}_i\cdot\vec{\nabla}W(\vec{x}_i) \sin m\varphi_i + m\dot{\varphi}_i W(\vec{x}_i) \cos m\varphi_i \right],
\end{align}
\end{subequations}
the Fourier amplitude, phase, and their time derivatives are
\begin{subequations}
	\label{eqs:Fourier}
\begin{align}
    \label{eq:Sigma:m}
	\Sigma_m 		&= \sqrt{C_m^2 + S_m^2},&
	\frac{\dot{\Sigma}_m}{\Sigma_m} &= \frac{C_m\dot{C}_m + S_m\dot{S}_m}{C_m^2+S_m^2},\\
	\psi_m 		 	&= \frac1m \tan^{-1} \frac{S_m}{C_m},&
	\Omega_m 	&= \frac{C_m\dot{S}_m - S_m\dot{C}_m}{m(C_m^2+S_m^2)}.
\end{align}
\end{subequations}
For the correct $\Sigma_m$, the window function must be normalised to $N_W \equiv 2\upi \int W R\diff R=1$ (or $\Sigma_m$ from equation~\ref{eq:Sigma:m} divided by $N_W$).

Each of the terms in equations~\eqref{eq:Cm,Sm} and~\eqref{eqs:dt:Cm,Sm} is of the form $\sum_i p_i = N \langle p\rangle$, i.e.\ is a sample mean. Hence, its variance can be estimated as
\begin{align}
	\sigma^2 \approx \frac{N}{N-1} \sum_i \big(p_i-\langle p\rangle\big)^2,
\end{align}
and equivalently the co-variances between any two such quantities. From these, the co-variance matrix for the derived quantities in equations~\eqref{eqs:Fourier} can be estimated via linear error propagation.

\section{Identifying the bar region}
\label{app:details}
We again assume that all positions and velocities are relative to the centre (determined before) and rotation is around the $z$-axis.

Our implementation first sorts particles in cylindrical radius $R$ and assigns a number $N_1$ of radial bins in $R$ (annuli). The innermost bin starts at $R_{0,\min}=0$ and extends to $R_{0,\max}$ such that $N_{\min}$ particles are contained. Each subsequent bin starts at $R_{i,\min}$ equal to the next particle just outside the previous bin and contains at least $N_{\min}$ particles, but more if $R_{i,\max}/R_{i,\min} < 10^{\Delta}$ with parameter $\Delta=0.15$ by default, though not exceeding a certain maximum $N_{\max}$. Next, we add $N_1-1$ intermittent bins that cover the radii between the medians of two adjacent primary bins. This gives $N_{\mathrm{a}}=2N_1-1$ overlapping cylindrical bins. In each of these we perform the $m=2$ Fourier analysis of Appendix~\ref{app:Omega} to determine $A_2=\Sigma_{m=2}/\Sigma_{m=0}$ and the phase $\psi$. For this purpose, we employ the top-hat window, which is more efficient, slightly more accurate (the effective particle number is higher than with non-uniform weighting), and sufficient for unbiased estimates of $\Sigma_{m}$ and $\psi_{m}$.

If the maximum $A_2$ across all bins is below a threshold (0.2 by default), we do not attempt to identify a bar. Otherwise, we start by setting the bar region to the radial bin with maximum $A_2$ and extend it as follows. We consider the next inner and outer bins for extension, if their $A_2$ exceeds half the maximum. If both qualify, we take that which keeps the range $\Delta\psi$ of phases covered by the bar region smallest. In this way, the bar region is extended until $A_2$ of the candidate bins is too small or $\Delta\psi$ would exceed a certain width (typically $10^\circ$). The inner and outer edge of the bar region are then identified as $R_0$ and $R_1$ of, respectively, the inner- and outermost bin in the bar region.

\label{lastpage}
\end{document}